# Bragg guiding of domain-like nonlinear modes and kink arrays in lower-index core structures

Fangwei Ye, Yaroslav V. Kartashov, Victor A. Vysloukh, and Lluis Torner

*ICFO-Institut de Ciencies Fotoniques, and Universitat Politecnica de Catalunya,*

*Mediterranean Technology Park, 08860 Castelldefels (Barcelona), Spain*

We introduce a novel class of stable nonlinear modes trapped in a lower-index film core sandwiched between two optical lattices, or in the cylindrical core of a radial lattice, imprinted in defocusing media. Such family of nonlinear modes transforms into defect lattice solitons when the core width is sufficiently small or into an array of kinks when the width is large enough. We find that higher-order modes with multiple zeros inside the guiding core can be stable in one-dimensional settings.

*OCIS codes: 190.0190, 190.6135*

Linear photonic bandgap structures, such as PCFs, with a high-contrast refractive index modulation have attracted broad interest during the last decade [1-4]. Low-contrast optically induced nonlinear PCF-like structures were introduced too [5]. On the other hand, a single interface between periodic and uniform media can guide light, both in focusing and defocusing case [6-12]. Lattice surface solitons are either associated with the semi-infinite photonic gap induced by total internal reflection [6,7] or with finite photonic gaps induced by Bragg reflections [8-15]. The latter includes surface gap solitons at defocusing lattice interfaces [10,11], and hybrid mixed-gap states at interface of different lattices [12]. The surface of a semi-infinite defocusing lattice also supports kink solitons [13]. Formation of optical domain walls is possible in quadratic nonlinear waveguide arrays [16]. A challenging open problem is to link nonlinear modes of PCF-like structures and surface waves.

In this Letter we address nonlinear modes of a planar low-index core sandwiched between two harmonic lattices imprinted in defocusing cubic Kerr media, as well as its radially symmetric two-dimensional analog. The nonlinear modes of such structures ex-hibit flat-top profiles in the core region and oscillatory decaying wings inside the lattices. By varying the core width these modes transform into defect solitons [17,18] in the small core limit and into a pair (for fundamental mode) or an array (for higher-order modes) of kinks [13] in the large core limit. In practice flat-top beams are potentially interesting, e.g., for designing optical limiters. By analogy with solid-state physics we term such flat-top modes with well-defined borders and abruptly decaying wings as domain solitons.

We first consider light trapping in the planar uniform core region between two lattices in a defocusing cubic medium, described by the nonlinear Schrödinger equation for the dimensionless complex amplitude of the light field $q$:

$$i\frac{\partial q}{\partial \xi} = -\frac{1}{2}\frac{\partial^2 q}{\partial \eta^2} + |q|^2 q - pR(\eta)q. \qquad (1)$$

Here the transverse $\eta$ and longitudinal $\xi$ coordinates are scaled to the characteristic scale $x_0$ and diffraction length, respectively; the parameter $p$ characterizes the lattice depth, while the function $R(\eta)$ describes the refractive index profile: $R(\eta) = \cos^2(\Omega\eta)$ for $|\eta| \geq n\pi/2\Omega$ and $R(\eta) = 0$ otherwise, $\Omega$ is the frequency and $n$ is the number of lattice periods removed to form the guiding core.

To understand the basic properties of the guided modes, it is instructive to consider the Floquet-Bloch spectrum of the infinite lattice. The band structure of the linear version of Eq. (1) with $R(\eta) = \cos^2(\Omega\eta)$ for $\eta \in (-\infty, +\infty)$ at $\Omega = 2$ is shown in Fig. 1(a). For each lattice depth all possible propagation constants $b$ of the Bloch modes form bands. Because of the defocusing nonlinearity, localized solutions can only be found in the finite gaps. One searches for the solution of Eq. (1) associated with the first finite gap in the form $q(\eta, \xi) = w(\eta)\exp(ib\xi)$. The fundamental mode exhibits a flat-top profile inside the core and decaying oscillating wings in the lattices; for thick-core structures ($n \gg 1$) it resembles a symmetric pair of kinks [Fig. 1(b)]. The light is trapped due to a balance between diffraction (enhanced by defocusing nonlinearity) and Bragg-type reflection from many periods of the lattices surrounding the core.

The localization of nonlinear modes depends on the position of $b$ inside the existence region $b_{\text{low}} \leq b \leq b_{\text{upp}}$ [Fig. 1(c)]. Nonlinear modes are well localized in the middle of the existence region, and their wings decay rapidly inside the lattices. When $b$ approaches the lower cutoff, $b_{\text{low}}$, the mode wings strongly penetrate into the bulk of the lattice, and the energy flow $U = \int_{-\infty}^{\infty} |q|^2 d\eta$ carried by the mode grows [Fig. 1(d)]. The lower cutoff coincides with the lower gap edge and it is independent of the core width. The existence region remarkably expands when the core width grows [Fig. 1(c)]. This is accompanied by an increase of the peak amplitude (which approaches $|b|^{1/2}$ for sufficiently big $n$), and by a dramatic growth of $U$ [Fig. 1(d)]. Importantly the region of existence approaches the one that corresponds to surface kinks at $n \to \infty$ [13]. Thus the family of fundamental modes bridges the gap between localized defect solitons at small $n$ and symmetric kink pairs at large $n$. Kink pairs exists only for lattice depths below a certain maximal value [Fig. 1(c)]. The steepness of kinks increases with diminishing $b$.

If the right lattice is different from the left one, then their band structures are relatively shifted. When such shift is large enough, an overlap between gaps of different orders is possible, which may lead to the emergence of mixed-gap solutions. The latter refers to hybrid modes whose propagation constants reside, for instance, in the first finite gap of the left lattice and the second finite gap of the right lattice. For example, at $p = 14$ if the right lattice has the frequency $\Omega_1 = 2$ while the left one $\Omega_2 = 5$, the second finite gap of the right lattice $b \in (-2.9, -0.49)$ overlaps with the first finite gap of left lattice $b \in (-8.9, -1.9)$ in the region $b \in (-2.9, -1.9)$. The mixed-gap nonlinear mode may emerge from this overlap region [Fig. 1(e)]. The existence region of the mixed-gap solutions is shown in Fig. 1(f); it shrinks when the difference between lattice frequencies grows.

Higher-order modes (anti-symmetric dipole one [Fig. 1(b)], tripole, etc) also were found numerically, and in the wide-core limit their profiles mimic the corresponding arrays of kink waves. The stability of higher-order modes were examined by a linear stability analysis and by direct integration of Eq. (1) in the presence of random perturbations (input conditions $q \mid_{\xi=0} = w(1 + \rho)$, where $\rho(\eta)$ stands for broadband noise with the variance $\sigma_{\text{noise}}^2 = 0.01$). Surprisingly, the fundamental and higher-order modes were found to be stable in most part of their existence region. Illustrative examples of stable propagation of fundamental, dipole and mixed-gap modes are shown in Figs. 2(a)-2(c). Only for $b \to b_{\text{low}}$ one encounters week oscillatory instability that may lead to pronounced radiation [Fig. 2(d)].

The above results correspond to one-dimensional geometries, but on physical grounds one expects them to be more general. In particular, two-dimensional radially symmetric structures featuring lower-index cores may also support similar kink-like (or domain) modes. Light guiding in such structures is governed by the two-dimensional version of Eq. (1) with a lattice having the profile $R = 0$ for $r < r_{\text{out}}$ and $R = \cos^2(\Omega r)$ for $r \geq r_{\text{out}}$, where $r^2 = \eta^2 + \zeta^2$ is the radial distance, $r_{\text{out}} = (2n-1)\pi/2\Omega$ is the core radius, $\Omega$ is the frequency, and $n$ sets the number of rings removed from the lattice to form the core. Fundamental radially symmetric nonlinear modes were found in the form $q(r,\phi,\xi) = w(r)\exp(ib\xi)$. Such modes also feature a flat plateau inside the wide lattice core and pronounced decaying oscillations in the lattice depth [Fig. 3(a)]. These oscillations are especially evident close to cutoffs $b_{\text{low}}$ and $b_{\text{upp}}$ (at small enough $p$), which clearly indicates that trapping is achieved due to the Bragg-type reflection from the radially periodic structure. Notice that linear light guiding in photoinduced ring lattices was demonstrated in [19], while solitons in periodic radial structures were found in [20].

One can see from Fig. 3(a) how at a fixed value of $b$, the mode, being well localized for small $n$, shows a flat-top domain-like shape with growth of the core radius. Analogously to one-dimensional case, the lower cutoff of existence region is independent of the core radius. For sufficiently high $p$, the mode amplitude and energy flow monotonically decrease with $b$ and vanish at the upper cutoff [Fig. 3(c)], while solutions remain well localized. However when $p$ is sufficiently small, the field penetrates deeply into the lattice even in the vicinity of $b_{\text{upp}}$, causing an abrupt divergence of the corresponding energy flow at $b \to b_{\text{upp}}$ [Fig. 3(c)]. In this regime $b_{\text{upp}}$ rapidly reduces with decreasing $p$.

The region of nonlinear mode existence expands when the core radius grows [Fig. 3(b)]. Interestingly, the existence region at $n \to \infty$ also approaches that of the one-dimensional mode in lattices with frequency $\Omega$ (since at $r \to \infty$ the term $(1/r)d/dr$ in Laplacian can be neglected and one gets effectively one-dimensional periodic system). Note the possibility to vary the guided power in a wide-range by changing the number of lattice periods.

The linear stability analysis revealed that two-dimensional nonlinear modes are stable in most part of their existence region, unless $b$ is too close to $b_{\text{low}}$ where we found weak oscillatory instabilities (with real parts of the perturbation growth rate $\delta_r \sim 10^{-4}$), which are reminiscent of the instabilities of usual gap solitons arising due to inter-band coupling. A decay scenario for such two-dimensional modes is similar to that shown in Fig. 2(d) for one-dimensional case. A small region of exponential instability also appears in shallow lattices close to the upper cutoff where $dU/db \geq 0$ [Fig. 3(d)]. Numerically, we verified the robustness of stable domain modes by propagating them over indefinitely long distances even in the presence of strong input perturbations [Figs. 2(e) and 2(f)]. In contrast to one-dimensional case we found that all higher-order solutions featuring radial or tangential nodal lines inside the core tend to be unstable. Both one- and two-dimensional domain-like modes could be excited with Gaussian or super-Gaussian beams with properly selected widths and peak intensities.

Summarizing, we introduced a novel type of nonlinear modes trapped in lower-index core between two one-dimensional defocusing lattices or in cylindrical guiding core of radially symmetric lattice. The domain-like nonlinear modes transform into defect lattice solitons when the lattice core is sufficiently narrow or into an array of surface kinks when the lower-index core width is large. Higher-order domain-like modes are stable in one-dimensional settings, but unstable in two-dimensional guiding structures.

# References with titles

# References without titles

# Figure captions

Figure 1.  (a) Band-gap lattice spectrum at $\Omega=2$. (b) Profiles of fundamental (black line) and dipole (red line) nonlinear modes at $b=-0.6$, $p=4$, $n=7$, $\Omega=2$. (c) Existence regions at $(p,b)$ plane for $n=7$ (black line) and $n=1$ (red line) at $\Omega=2$. (d) Energy flow versus $b$ for $p=4$, $\Omega=2$. (e) Mixed-gap modes at $b=-2$ (1) and $b=-2.7$ (2) for $p=14$, $\Omega_1=2$, $\Omega_2=5$. (f) Existence regions at $\Omega_1=2$ for $\Omega_2=4$ (black line) and $\Omega_2=5$ (red line). In (e) and (f) the core width is $4\pi/\Omega_1+4\pi/\Omega_2$. Green lines in (c) and (f) show gap edges.

Figure 2.  Stable propagation of fundamental (a) and dipole (b) one-dimensional modes at $b=-0.6$, $p=4$, $n=7$, $\Omega=2$ and (c) mixed-gap mode at $b=-1$, $p=20$, $\Omega_1=2$, $\Omega_2=4$. (d) Dynamics of unstable fundamental mode at $b=-1.44$, $p=2$, $n=7$, $\Omega=2$. Stable propagation of two-dimensional domain soliton in lattice with $n=5$ and $\Omega=2$ at $b=-0.7$, $p=4$. In (e) and (f) field modulus distributions are shown at different distances.

Figure 3.  (a) Fundamental mode profiles in circular lattices at $b=-0.6$ and $p=4$. (b) Existence region on the $(p,b)$ plane for $n=2$ (black line) and $n=5$ (red line). (c) Energy flow versus $b$ at $n=5$. (d) Real part of perturbation growth rate versus $b$ at $p=1.5$, $n=5$. In all cases $\Omega=2$.

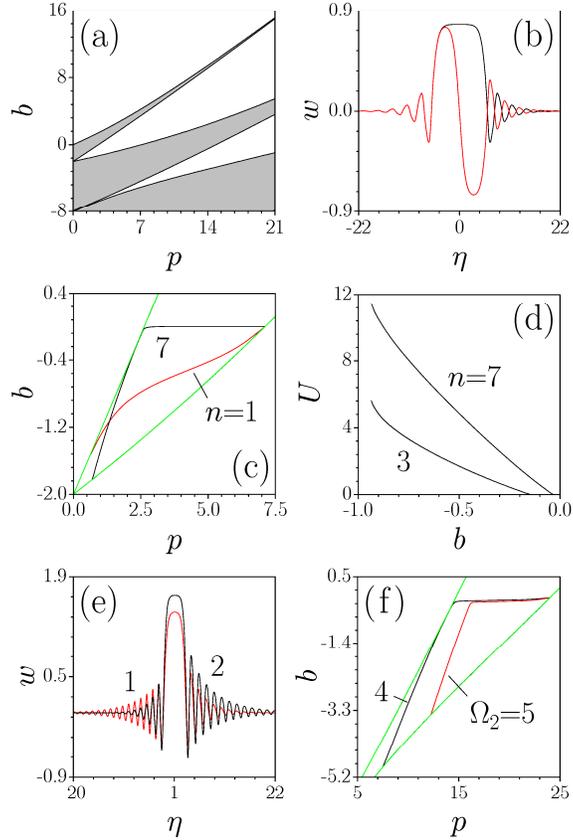

Figure 1. (a) Band-gap lattice spectrum at $\Omega = 2$. (b) Profiles of fundamental (black line) and dipole (red line) domain solitons at $b = -0.6$, $p = 4$, $n = 7$, $\Omega = 2$. (c) Regions of soliton existence on $(p,b)$ plane for $n = 7$ (black line) and $n = 1$ (red line) at $\Omega = 2$. (d) Energy flow versus $b$ for $p = 4$, $\Omega = 2$. (e) Mixed-gap domain solitons at $b = -2$ (1) and $b = -2.7$ (2) for $p = 14$, $\Omega_1 = 2$, $\Omega_2 = 5$. (f) Regions of soliton existence at $\Omega_1 = 2$ for $\Omega_2 = 4$ (black line) and $\Omega_2 = 5$ (red line). In (e) and (f) the core width is $4\pi/\Omega_1 + 4\pi/\Omega_2$. Green lines in (c) and (f) show gap edges.

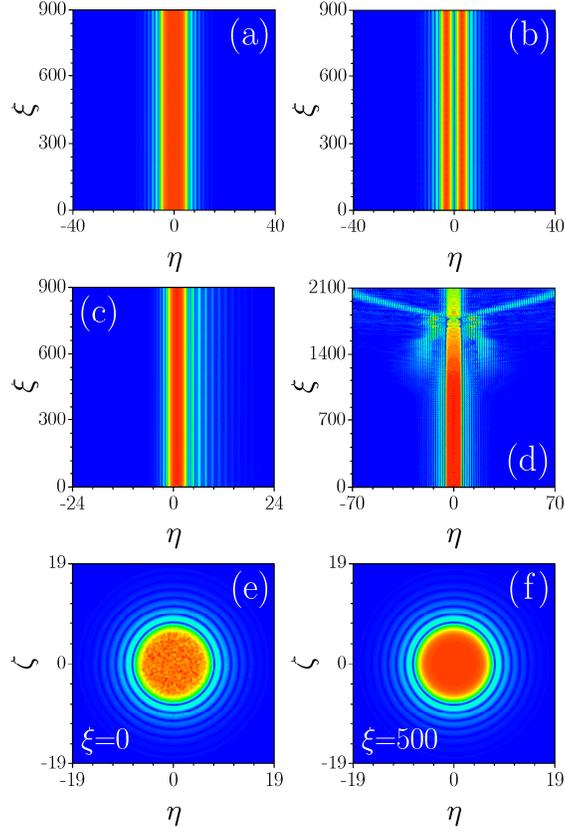

Figure 2. Stable propagation of fundamental (a) and dipole (b) one-dimensional modes at $b = -0.6$, $p = 4$, $n = 7$, $\Omega = 2$ and (c) mixed-gap mode at $b = -1$, $p = 20$, $\Omega_1 = 2$, $\Omega_2 = 4$. (d) Dynamics of unstable fundamental mode at $b = -1.44$, $p = 2$, $n = 7$, $\Omega = 2$. Stable propagation of two-dimensional domain soliton in lattice with $n = 5$ and $\Omega = 2$ at $b = -0.7$, $p = 4$. In (e) and (f) field modulus distributions are shown at different distances.

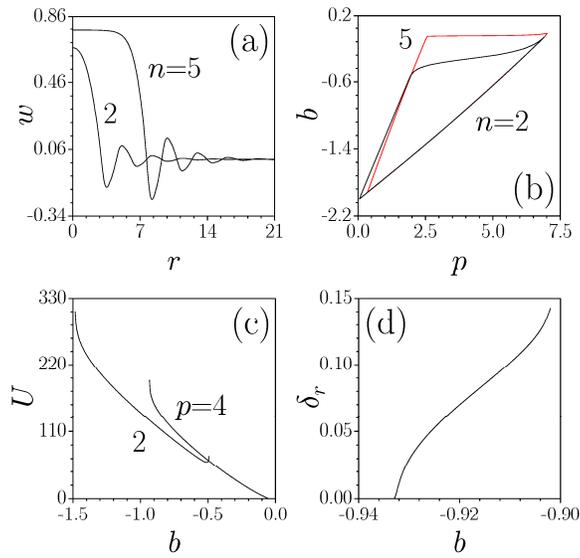

Figure 3. (a) Soliton profiles in circular lattices at $b = -0.6$ and $p = 4$. (b) Region of soliton existence on the $(p,b)$ plane for $n = 2$ (black line) and $n = 5$ (red line). (c) Energy flow versus $b$ at $n = 5$. (d) Real part of perturbation growth rate versus $b$ at $p = 1.5$, $n = 5$. In all cases $\Omega = 2$.

12